\documentclass{PoS}
\usepackage{booktabs,subcaption,amsfonts,dcolumn}
 \usepackage{multicol}
\usepackage{floatrow}
\usepackage{wrapfig}
\newfloatcommand{capbtabbox}{table}[][\FBwidth]

\usepackage{blindtext}
\title{Effects of the atmospheric electric field on the HAWC scaler rate.}

\ShortTitle{Effects of the atmospheric electric field on the HAWC scaler rate.}

\author{\speaker{Angel Ricardo Jara Jimenez}\\
        Universidad Veracruzana, M\'exico.\\ %for the HAWC collaboration\footnote{for collaboration list see PoS(ICRC2019)1177}\\
        E-mail: \email{rt\_jaja@outlook.com}}

\author{Alejandro Lara\\
Instituto de Geof\'isica-UNAM, M\'exico\\
        The Catholic University of America, Washington DC, USA.\\ %for the HAWC collaboration\footnote{for collaboration list see PoS(ICRC2019)1177}\\
        E-mail: \email{alara@igeofisica.unam.mx}}

\author{Arun Babu Kollamparambil Paul\\
        Instituto de Geof\'isica-UNAM, M\'exico.\\ %for the HAWC collaboration\footnote{for collaboration list see PoS(ICRC2019)1177}\\
        E-mail: \email{arun@igeofisica.unam.mx}}
     
\author{ James Ryan\\
        University of New Hampshire, USA. \\
        E-mail: \email{jimunhryan@me.com}}
\author{for the HAWC collaboration\footnote{for collaboration list see PoS(ICRC2019)1177}}

\abstract{Strong electric fields in thunderclouds have long been known to accelerate secondary charged particles. We investigate this effect using three years (2015-2017) of data from the scalar system of the High Altitude Water Cherenkov (HAWC) observatory, which is an air shower array deployed 4100 m a.s.l. in central Mexico. The experimental site is frequently affected by strong thunderstorms, and the detector's high altitude, large area, and high sensitivity to cosmic-ray air showers make it ideal for investigating particle acceleration due to the electric fields present inside the thunder storm clouds. In particular, the scaler system of HAWC records the output of each one of the 1200 PMTs as well as the 2, 3, and 4-fold multiplicities (logic AND in a time window of 30 ns) of each water Cherenkov detectors (WCD) with a sampling rate of 40 Hz. Using data from this scaler system, we identify approximately 100 increases in the scaler rate which is in time coincidence with thunderstorms. These events show high cross correlation between the scaler rate and the electric field, hence can be produced by the acceleration of secondary particle by the thunderstorm electric fields. In this work we present the method of identification of these events and their general characteristics.}

\FullConference{36th International Cosmic Ray Conference -ICRC2019-\\
		July 24th - August 1st, 2019\\
		Madison, WI, U.S.A.}

\begin{document}

\section{Introduction}
It is well known that strong electric field generates large enough
forces on electrons to overcome the drag forces and also can remove secondary electrons from these atoms, these are
the so called run-away electrons.(\cite{1478-7814-37-1-314},
\cite{doi:10.1029/2006GL025863}). A secondary electron, produced by  
cosmic-rays or radioactive decay (the seed particle) with enough
energy (hundreds of keV ) accelerated by a strong electric field may
be able to produce at least one run-away electron in the range of tens
to hundreds keV, which in turn can create additional run-away
electrons, producing an avalanche multiplication. In this way, each
energetic seed electron injected into a high electric field region will result in a large number of
relativistic run-away electrons. This mechanism is known as Relativistic Runaway Electron
Avalanche (RREA). The flux of relativistic run-away electrons experiences a deceleration due to
the interaction with atoms and ions, producing high energy photons by bremsstrahlung emission, (\cite{article}, \cite{article8}).\\
These events have beeen detected by satellite gamma ray detectors and
are known as terrestrial gamma ray flashes,
(\cite{10.2307/2884079}). At ground  level , high altitude cosmic ray
detectors have reported ground enhancements during thunderstorm,
(\cite{article2},
\cite{10.1029/JZ072i018p04623},\cite{PhysRevD.82.043009}). The HAWC observatory has also report this kind of enhancements, \cite{article9}. We report on a new method to find increases in the count rate in HAWC dataset, and select those that are due to cloud electric fields by finding the correlation with the electric field intensity.  We show the results corresponding to the years of 2015-2017.\\

\section{HAWC scaler system}
The High Altitude Water Cherenkov (HAWC) observatory is located on a relatively flat piece of land near the saddle region between the Sierra Negra and Pico de Orizaba, with lattitude $18^{\circ}59' 41"$ N,  longitude $97^{\circ} 18'30.6"$ W and altitude at 4100 m above sea level. 
HAWC consist of 300 water Cherenkov detectors (WCD),  each of them are of 7.3 m in diameter and 5.0 m in depth. They are spread over an area of 20,000 $m^2$. Each of these WCDs is filled with filtered water and instrumented with  4 photomultiplier tubes (PMTs) a 10 inch PMT at the center of the WCD at positions $'C'$ and three 8-inch PMTs being arranged around the central one, $C$, making an equilateral triangle of side 3.2 m at positions $A$, $B$ and $D$.
\\
HAWC has two different kinds of Data Adquisition System(DAQ).The main DAQ measures arrival times and time over thresholds of PMT pulses. It also has a TDC scaler system which counts the hits within a time window of 30 ns of each PMT and the coincidences of 2, 3 and 4 in each water Cherenkov detector. These coincidences are called multiplicity 2, 3 and 4, respectively. The secondary DAQ consists of a counting system that registers each time the PMT is hit by $\>$ 1/4 photoelectron charge and it is called hardware (HW) scaler system. The seconadary DAQ is sensitive to gamma-rays, cosmic-rays, solar activity and transitive events that increase or decrease the count of rates.\\ The response of each PMT is sent to a control room where is digitized by time to digital converters (TDC) so it can be read by a computer (\cite{2018NIMPA.888..138A}, \cite{2012APh....35..641A}, \cite{article4}).

\section{Data Selection}
\subsection{HAWC Data}
\begin{figure}[h]
\center
\includegraphics[width=.8\textwidth]{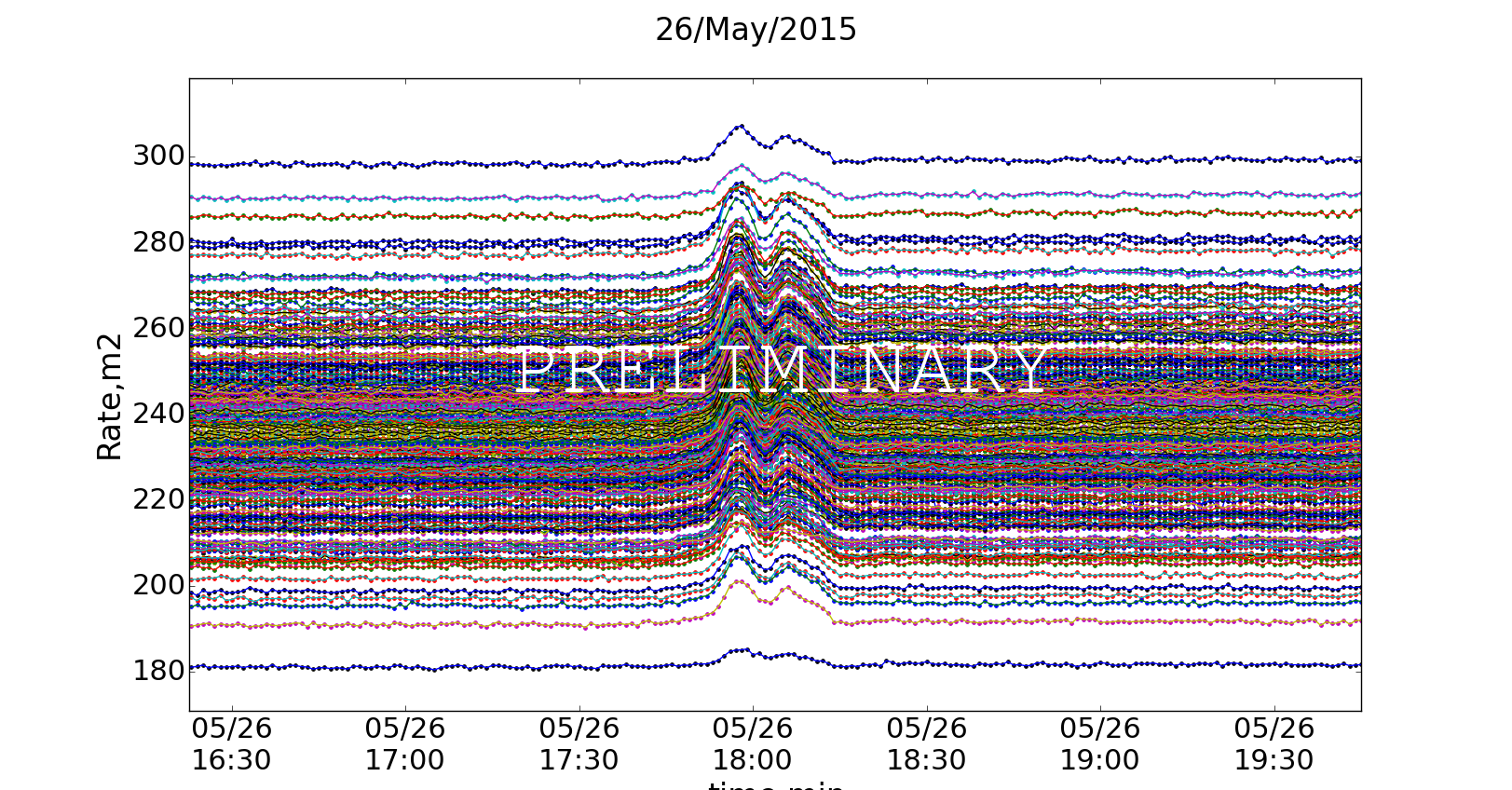}
 \caption{26/May/2015, multiplicity 2, each curve correspond to a detector.} 
\label{original}
\end{figure}

The time period covered by  this work corresponds to  2015-2017 and the data recording has a time cadence of one minute to improve the stadistical accuracy of the measurements. In our study, we analyze the information of each multiplicity (2,3 and 4) of the 300 detectors. The electric field data used for the study was measured in kV/m and have in general a resolution of 1 minute. During the three years there are some exceptions that will be explained in section 3.2.2\\
The data selection was made using three-step algorithm that we call filters. This filters only select the data that fullfill the specified requirements. To have a better and clear idea of how the study was made, we will use as example a event where the electric field was not saturated and was observed by the two scaler systems, the event of 2015 May 26. In Fig.~\ref{original} the data for multiplicity 2 are shown.
\subsection{First Filter}
\begin{figure}[h]
\center
\includegraphics[width=.8\textwidth]{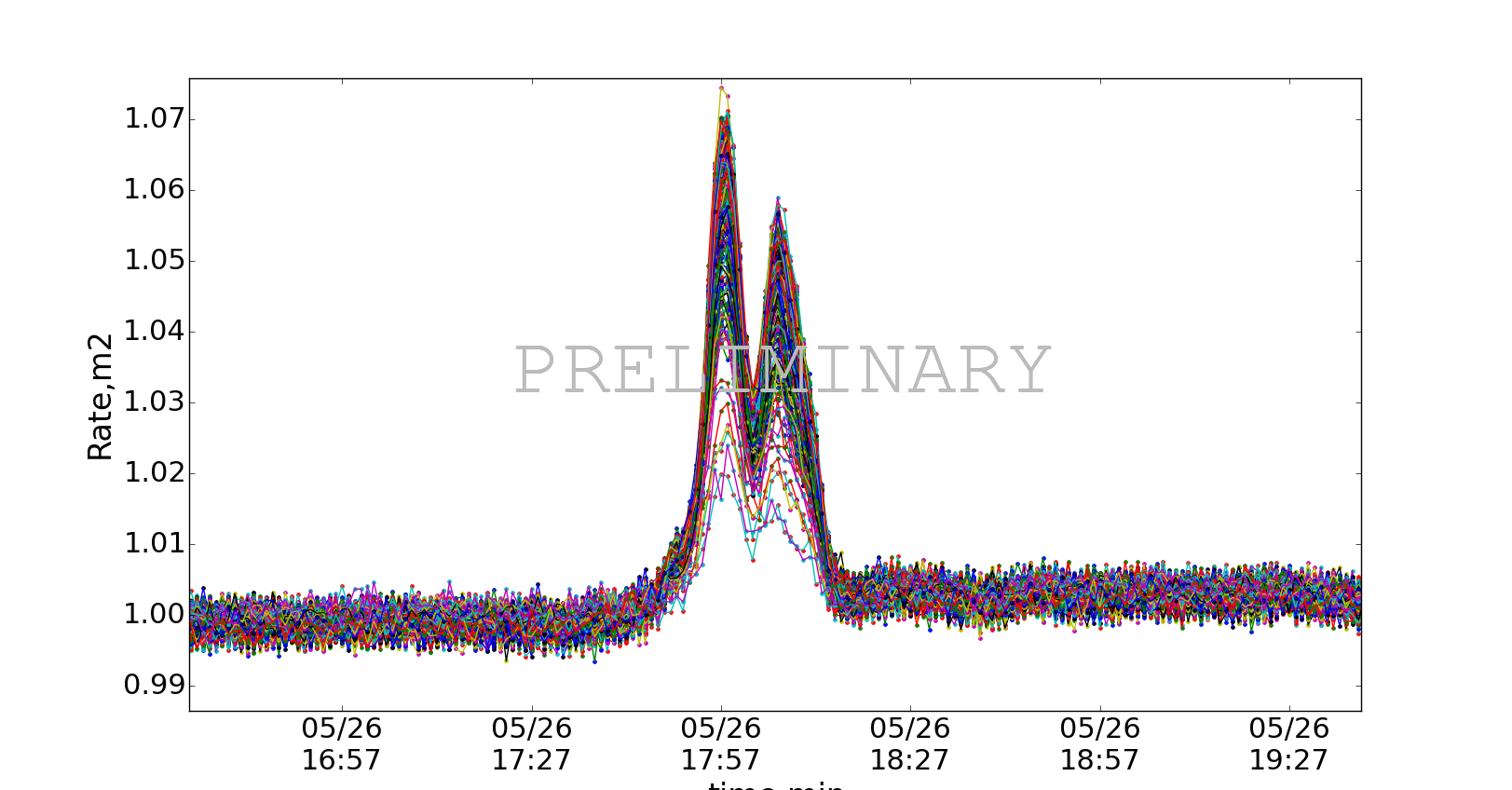}

 \caption{Example of the dataset, each color line represent a detector. The dataset correspond to the count of rate in multiplicity 2 for 2015 May 26. The y-axis shows the count rate as a fuction of time.}
\label{fig:examp}
\end{figure}

\begin{figure}[h]
\center
\includegraphics[width=.8\textwidth]{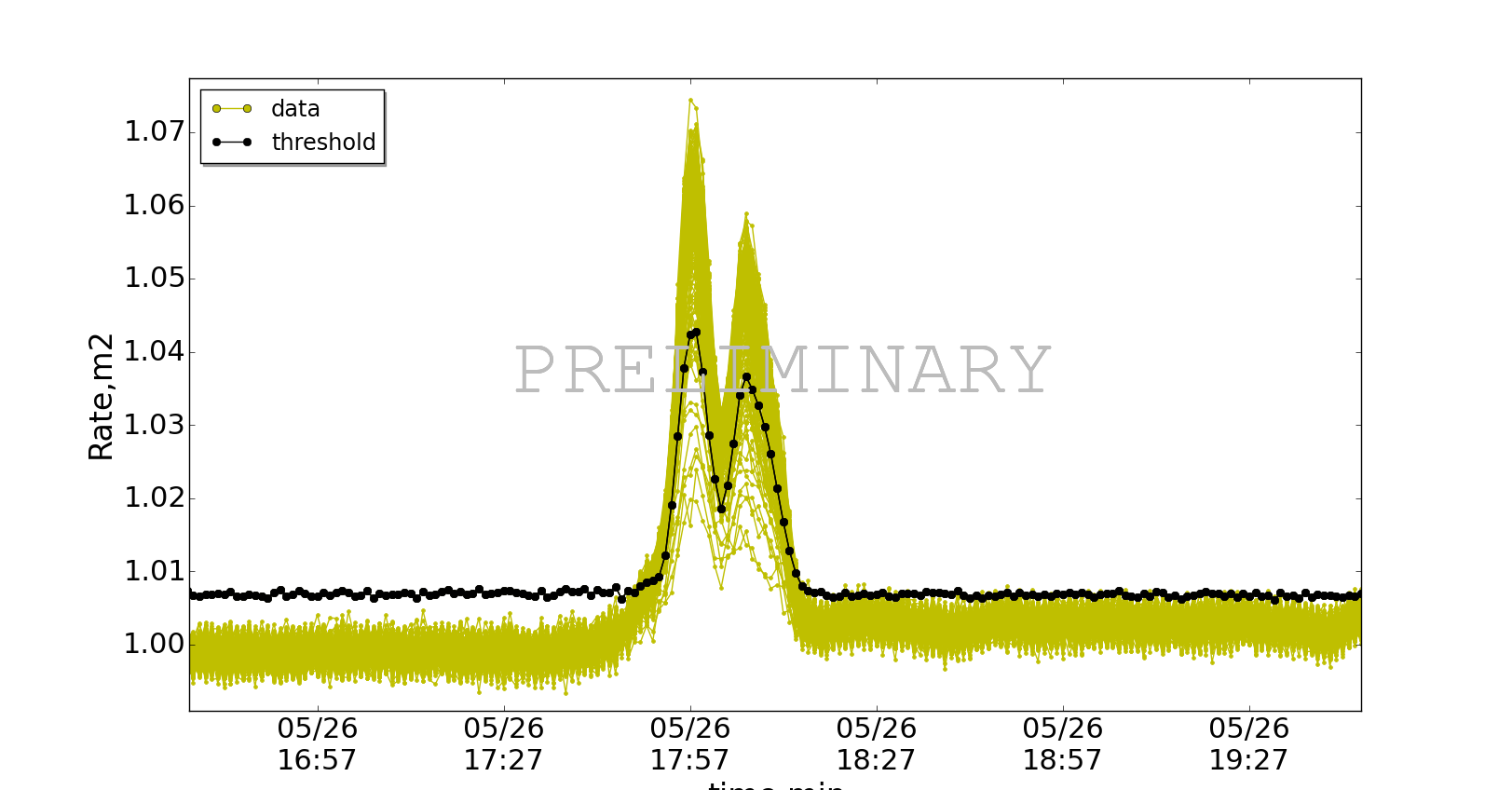}
 \caption{2015 May 26, multiplicity 2 data(yellow) and threshold(black)} 
\label{threshold}
\end{figure}

As mention above and shown in the example (Fig.~\ref{original}) each detector have his own average rate, so the first step was to obtain a standard average value for all the detectors. The standard average value was obtained by doing a selfnolmalization of each detector $i$ with a mean $p_i$. The normalized values, nv$_{i,t}$, were calculated as:
\begin{equation} 
 \label{eqn:1} 
nv_{i,t}=\frac{Sd_{i,t}}{p_i},
\end{equation}
where Sd$_{i,t}$ is the original value of the data at time $t$ corresponding to detector $i$. Fig.~\ref{fig:examp} shows what happens after selfnormalization.
%\subsubsection{First filter: Selecting Algorithm}
%\item Calculation of $\sigma$$_i$:
\\
Once the selfnormalization is done, the standard average value will be approximately 1, as seen in Fig.~\ref{fig:examp}. This will then be used to calculate a threshold which is then used to define the increments. With the standard deviation for each minute $t$, $\sigma_t$, the threshold is calculated according to Eq.~\ref{eqn:2}. For an increment to be deemed an event $nv_{i,t}>th_i$ must be true for at least 5 continuos minutes and registered in at least 30 detectors. These conditions were chosen so as to not select electronic malfunctions of a PMT. In Fig.~\ref{threshold} is graphic reperesentation of the count rate and the threshold. In this filter 202 events were found.
%\item Threshold(th)\\
\begin{equation}
\label{eqn:2}
th_i=1+5\sigma_i .
\end{equation}
\subsubsection{Second filter}
The duration of a thunderstorm cell is around 30  to 60 minutes \cite{feyma}. For that reason we choose the increments with a significant duration ($5 < t < 120$ minutes). The intervals longer than 2 hours are more likely to be due solar or atmospheric modulation, no the objective of this study. Only 100 intervals have the specified duration.
\subsubsection{Third filter}
The electric field detector, Boltek EFM-100, was installed on 2015 May, so the intervals before that date have not corresponding electric field data. There are also gaps because the detector could stop functioning from a nearby electric discharge or maintenance. Thus, in the 100 events only 79 had corresponding electric field data.
\section{Analysis}
For the subsequent steps in the analysis of the 79 events, we again use the event of 2015 May 26 as an example.
\subsection{Pressure correction}
A pressure correction was made to the selected events. In the method proposed by K.P. Arun Babu, the data are transformed as percentages, with the pressure data and the pressure coefficients for each multiplicity. Then the data are smoothed by using a moving central average with a window of 5 minutes, finally a linear transformation is made by setting the minimum value of the data equal to zero.
%$Spv_{i,t}$, by using the mean value of the background ($np_i$) as equation.~\ref{eqn:3} indicates. With the pressure data (ps$_t$) and its mean value (mp) we calculated pressure differential for each minute (dp$_t$), equation.~\ref{eqn:4}. The equation.~\ref{eqn:5} is used to calculate the corrected values, where $cm_{2,3 or 4}$ are the pressure coefficients for each multiplicity ontained by Arun Babu. 
%\begin{itemize}
 %   \item multiplicity 2 (cm$_2$) $\rightarrow$ 0.41$\frac{\%}{hPa}$
  %  \item multiplicity 3 (cm$_3$) $\rightarrow$ 0.31$\frac{\%}{hPa}$
   % \item multiplicity 4 (cm$_4$) $\rightarrow$ 0.25$\frac{\%}{hPa}$
%\end{itemize}
%\begin{equation}
%\label{eqn:3}
% Spv_{i,t}=\frac{nv_{i,t}-np_i}{np_i}100
%\end{equation}
%\begin{equation}
%\label{eqn:4}
 % dp_i =\mid ps_i -mp \mid 
%\end{equation}
%\begin{equation}
%\label{eqn:5}
 %        Hc_i= Hpv_ie^{-cm_{2,3 or 4}dp_{i}} 
%\end{equation} 
%The corrected data was smoothed by using a moving average with a window of 5 minutes and then a linear transformation was done to have a minimum value equal to zero. In Fig.~\ref{fig:prescoel} is shown the example dataset after the all the procedure explained in this section.
\subsection{Crosscorrelation}
\begin{figure}
\begin{floatrow}
\ffigbox{%
 \includegraphics[width=1.2\columnwidth]{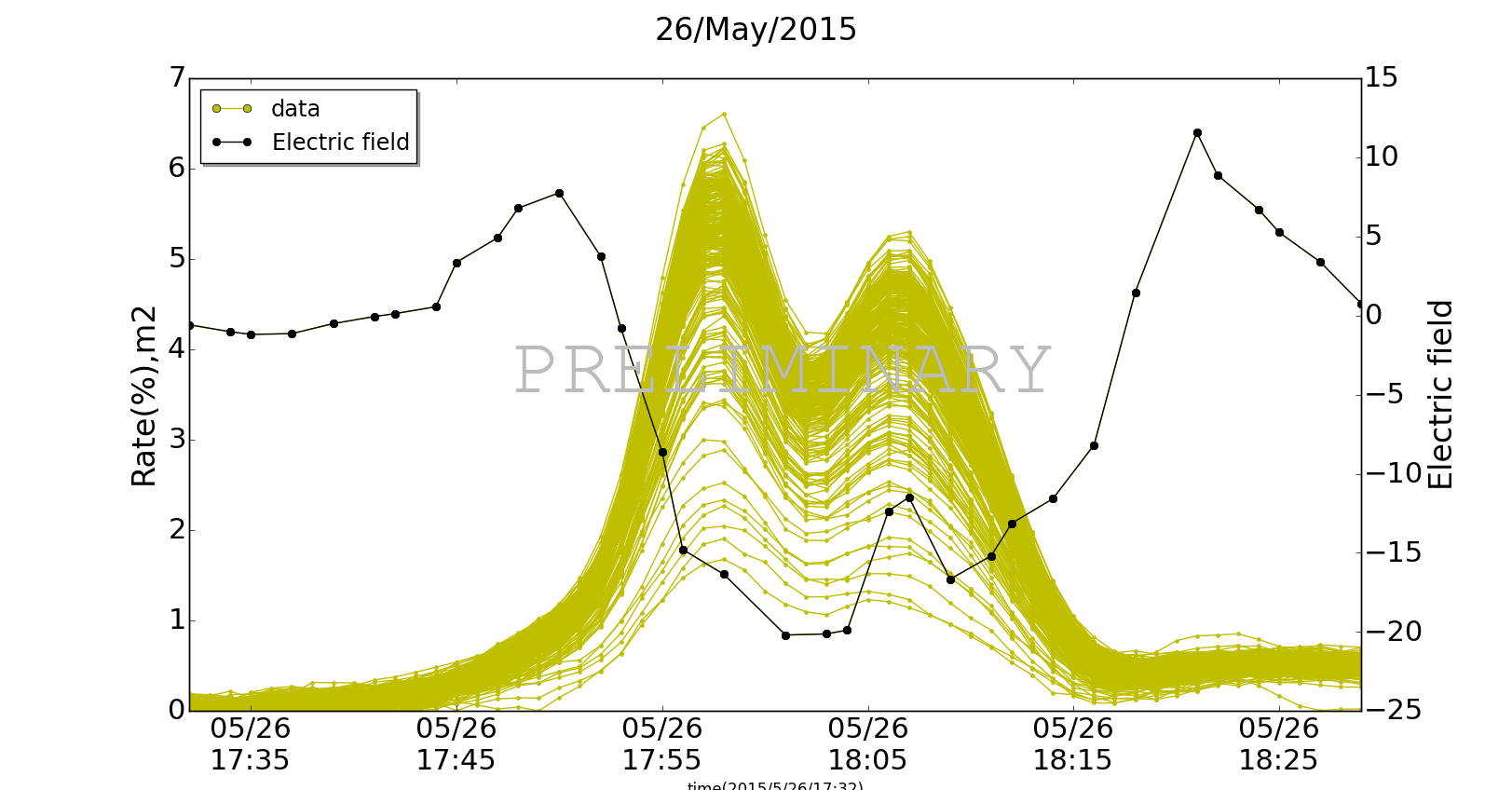}
}{%
 \caption{{\scriptsize  \it \label{fig:prescoel}26/May/2015,Comparison of the data with the electric field data.} }
}
\ffigbox{%
 \includegraphics[width = 1.2\columnwidth]{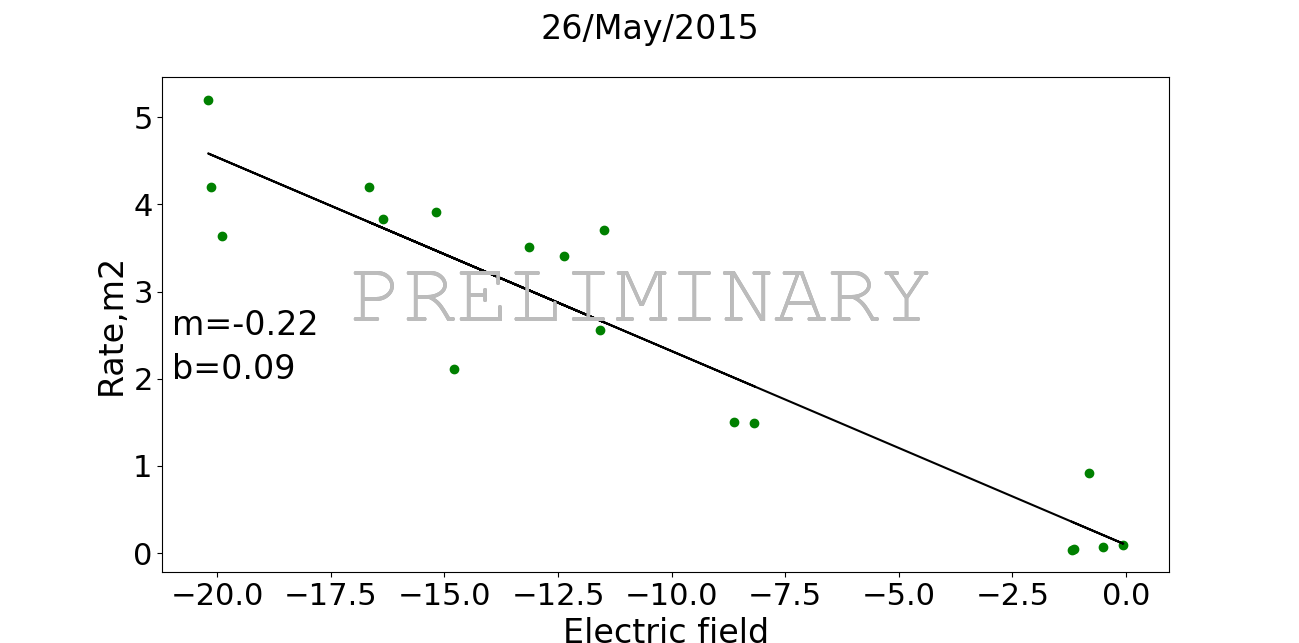}
}{%
\caption{{\scriptsize  \it \label{fig:fit} 26/May/2015,Linear fitting.  }}
}
\end{floatrow}
\end{figure}

\begin{figure}[h!]
 \center
\includegraphics[width=1.0\textwidth]{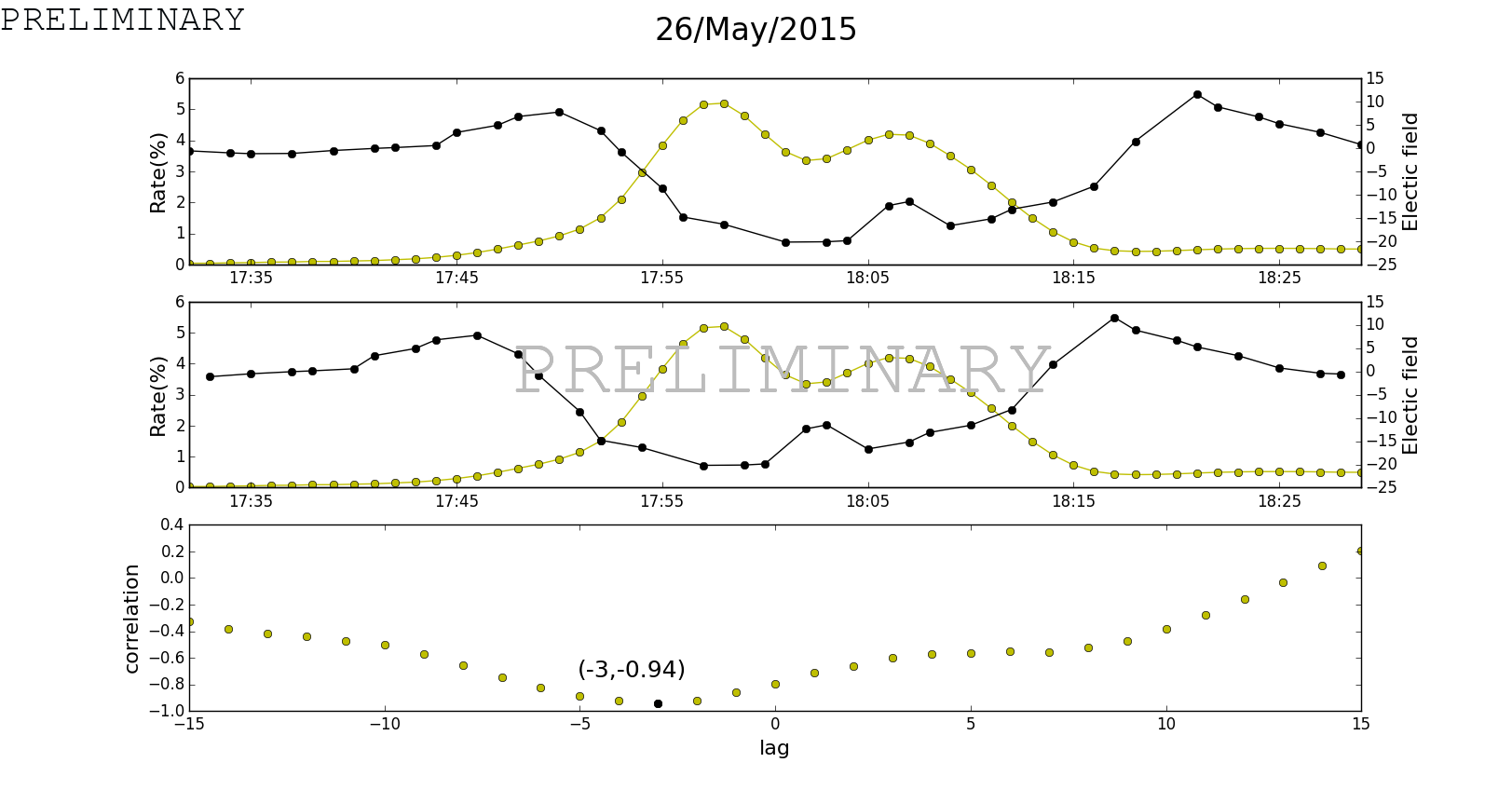}
 \caption{Up to down ,a)Comparison of the average rate with the electric field intensity.b) Comparison of the average rate with the shifted electric field. c)Cross correlation coefficient as a function of time lag.}  
\label{fig:avprescoel}
  \end{figure}

The electric field dependence of the 79 selected events was investigated first by comparing the count rate of the detectors and the corresponding electric field data, as in Fig.~\ref{fig:prescoel}. Subsequent analysis is carried out using the average rate of all the detectors, Fig.~\ref{fig:avprescoel}a.  
%\begin{figure}[h!]
 %\center
%\includegraphics[width=.8\textwidth]{preelecW.png}
 %\caption{26/May/2015,Comparison of the data with the electric field data.}  
%\label{fig:prescoel}
 % \end{figure}
The crosscorrelation is calculated by time shifting the electric field data by -15 to 15 minutes. The best shift for the electric field was obtain by calculating the crosscorrelation coefficient for each possible shift. In Fig.~\ref{fig:avprescoel}c we see correlation of the rates with the electric field for different offsets. In Fig.~\ref{fig:avprescoel}b we can see the comparison between the average data and the shifted electric field. Finally the dependecy of the electric field on the measure rate was estimated by using a linear fit only considering the negative values of the electric field, as in Fig.~\ref{fig:fit}. We can fit the dependecy of the electric field on the measure rate with a linear fit, where, in Eq.~\ref{eqn:he}, $R$ is the scaler rate and $E$, electric field intensity.
\begin{equation} 
 \label{eqn:he} 
R(E)=-0.22E+0.09
\end{equation}

%\begin{figure}[h!]
%\center
%\includegraphics[width=.8\textwidth]{fittW.png}
%\caption{26/May/2015,Linear fitting.} 
%\label{fig:fit}
%\end{figure}
\section{Discussion}
\begin{figure}[h!]
 \center
\includegraphics[width=0.9\textwidth]{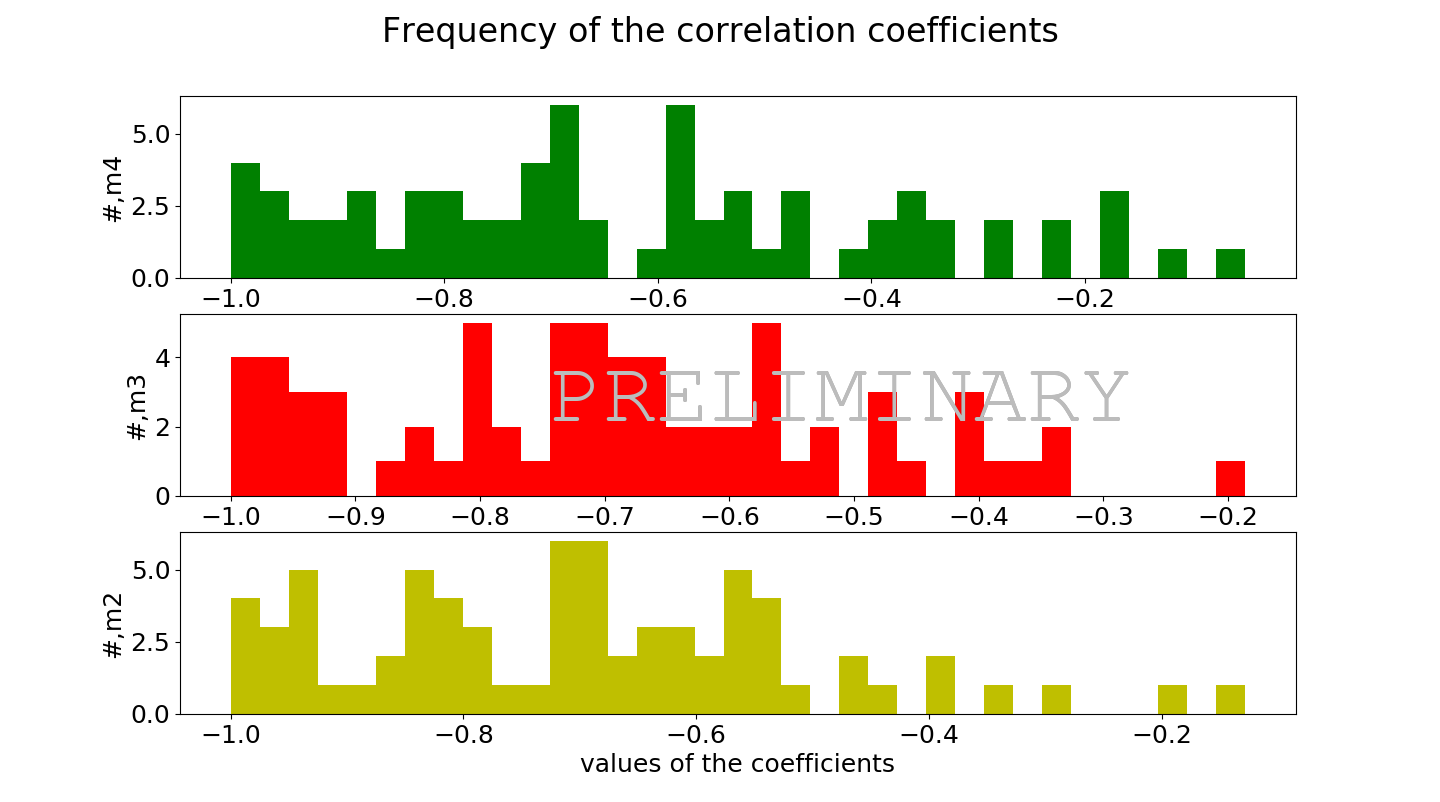}
 \caption{Histogram showing the frequency of the correlation coefficients.}  
\label{fig:hisco}
  \end{figure}

In this work , we propose a new method to identify intervals in the HAWC scalar data affected by atmospheric electricity activity. Of the 79 events (Tables.~\ref{tab_1},\ref{tab_2} and \ref{tab_2}), we see in Fig.~\ref{fig:hisco} that the majority of events show a strong dependence of the electric field intensity. These results are evidence for the idea that particle acceleration due to the electric field of clouds is producing an enhancements of the scaler rate of HAWC. After correcting for barometric pressure variations we see in Fig.~\ref{fig:hisco} that the majority of the events are inversely correlated with the electric field intensity. This implies that the enhancement in the scalar rates occurs almost concurrently with the decay in the electric field intensity. %In the case of the positive values of lag, the most probable scenario is that the enhancements is due a particle accelaration of clouds that at the beginning were far, but were moving towards the array, this can be veryfied by analysing the direction and speed of the wind in that moment.\\
 %The tendecy of the values of lag is shown in Fig.~\ref{fig:hislg}, where we observe the tendency of lag in the three multiplicities is around -7 and -6, also we observe a significant portion of positive values

%\begin{figure}[h!]
 %\center
%\includegraphics[width=1.0\textwidth]{histlgW.png}
 %\caption{Histogram showing the frequency of the lags.}  
%\label{fig:hislg}
 % \end{figure}
\begin{table}[h]
\begin{subtable}[t]{0.33\textwidth}
%\flushleft
\scriptsize
 %\centering
\begin{tabular}[t]{l c r}
\hline
\textbf{2015}&&\\
\hline
May,26& Jun,30b&   Sept,18\\
May,30& Jul,0910&  Sept,25,a\\
Jun,20& Jul,2021&  Sept,25,b\\
Jun,21& Aug,07&    Oct,28\\	
Jun,23& Aug,12,a&  Nov,02,a\\
Jun,30a& Aug,12,b& Nov,02,b\\
&             Aug,30&  \\
\hline
\end{tabular}
\caption{\footnotesize 2015}
\label{tab_1}
 \end{subtable}
%\end{table}
%\hspace{\fill}
\begin{subtable}[t]{0.33\textwidth}
%\flushright
\scriptsize
 %\centering
\begin{tabular}[t]{l c r}
\hline
\textbf{2016}&&\\
\hline
Jan,05,a&   Jun,01&   Jul,30\\
Jan,05,b&   Jun,05&   Aug,02\\
Mar,03&   Jun,08&     Aug,09,a\\
Mar,04&   Jun,17&     Aug,09,b\\
Apr,20&     Jun,1718& Aug,19\\
May,19&     Jun,20&   Aug,20\\
May,24&     Jun,27&   Sept,12\\
May,27,a&     Jun,30& Sept,17\\
May,27,b&     Jul,01& Sept,24\\
May,2728&   Jul,08&   Sept,29\\
May,28,a&   Jul,12&   Oct,01\\
May,28,b&   Jul,14&   Oct,16\\
\hline
\end{tabular}
\caption{\footnotesize 2016}
\label{tab_2}
\end{subtable}
%\bigskip 
%\end{table}
\begin{subtable}[t]{0.33\textwidth}
%\flushright
\scriptsize
 %\centering
\begin{tabular}[t]{l c r}
\hline
\textbf{2017}&&\\
\hline
Mar,16&    Jun,17&   Aug,10\\
Apr,07&   Jun,28&    Aug,11\\
Apr,08&   Jun,28&    Aug,15\\
Apr,19&   Jul,01&    Sept,14\\
May,04&   Jul,16,a&  Oct,10\\
May,13&   Jul,16,b&  Oct,11,a\\
May,1920& Aug,03&    Oct,11,b\\
Jun,07&   Aug,09&    Oct,11,c\\
\hline
\end{tabular}
\caption{\footnotesize 2017}
\label{tab_3}
\end{subtable}
\caption{Tables with the dates of the events}
\end{table}
If particle accelaration is occurring, it will possible to apply this method to other HAWC data to find more enhancements in all three multiplicities. From this,  it will be possible to calculate the energy of the incidents particles initiating the runaway. By considering this possibility we have shown the great utility of HAWC in the study of the mechanisms of particle acceleration
by the atmospheric electricity.\\
Further anlysis is necessary accompanied by simulations of the spatial and temporal response of HAWC. This will allow us to quantify the magnitude of the effect and the underlying physics. We will also perform an inverse analysis to verify the method, in this case the decay in the elctric field intensity will be determined first, from which we search for the corresponding enhacements in the rates. %This analysis will be published elsewhere.% A study of the wind direction and velocity for the events that present a positive value of lag is also necessary to polish the method.
\section{Acknowledgements}
We acknowledge the support from:
the US National Science Foundation (NSF);
the US Department of Energy Office of High-Energy Physics;
the Laboratory Directed Research and Development (LDRD) program of Los
Alamos National Laboratory;
Consejo Nacional de Ciencia y Tecnolog\'ia (CONACyT), M\'exico
(grants 271051, 232656, 260378, 179588, 254964,
258865, 243290, 132197)
(Cátedras 873, 1563, 341),
Laboratorio Nacional HAWC de rayos gamma;
L'OREAL Fellowship for Women in Science 2014;
Red HAWC, M\'exico;
DGAPA-UNAM (grants AG100317, IN111315, IN111716-3, IA102715, IN111419,
IA102019, IN112218),
VIEP-BUAP; PIFI 2012, 2013, PROFOCIE 2014, 2015;
the University of Wisconsin Alumni Research Foundation;
the Institute of Geophysics, Planetary Physics, and Signatures at Los
Alamos National Laboratory;
Polish Science Centre grant DEC-2014/13/B/ST9/945, DEC-2017/27/B/ST9/02272;
Coordinaci\'on de la Investigaci\'on Cient\'ifica de la Universidad Michoacana;
Royal Society - Newton Advanced Fellowship 180385.
Thanks to Scott Delay, Luciano D\'iaz and Eduardo Murrieta for technical
support.

\end{document}